\documentclass[11pt]{article}
\pdfoutput=1
\usepackage{float}
\usepackage{jheppub}
\usepackage{amssymb,amsfonts}
\usepackage{mathrsfs}
\usepackage{slashed}
\usepackage{graphics}
\let\savenumberline\numberline
\def\numberline#1{\savenumberline{#1.}}
\makeatletter
\renewcommand{\@seccntformat}[1]{\csname the#1\endcsname.\,\,}
\makeatother

\newcommand{\CB}{{\cal B}}

\newcommand{\CM}{{\cal M}}

\renewcommand{\hat}[1]{\widehat{#1}}
\newcommand{\be}{\begin{equation}}
\newcommand{\ee}{\end{equation}}
\newcommand{\bea}{\begin{eqnarray}}
\newcommand{\eea}{\end{eqnarray}}

\newcommand\secref[1]{{\S\ref{#1}}}

\newcommand\figref[1]{{Figure~\ref{#1}}}

\makeatletter
\def\@fpheader{\relax}
\makeatother
\usepackage{graphicx}
\usepackage{latexsym}

\title{\ \vspace{1.5in} \\ \hbox{Notes on the Quantization of Tropological Yang Mills Theory}}
\author{Emil Albrychiewicz, Andr\'{e}s Franco Valiente and Viola Zixin Zhao }
\affiliation{\medskip
Leinweber Institute for Theoretical Physics and Department of Physics\\
University of California, Berkeley, CA, 94720-7300, USA\medskip\\
Theoretical Physics Group, Lawrence Berkeley National Laboratory\\
Berkeley, CA 94720-8162, USA}

\emailAdd{ealbrych@berkeley.edu}
\emailAdd{andresfranco@berkeley.edu}
\emailAdd{zhaozixin@berkeley.edu}
\abstract{We continue our investigation into anisotropic topological field theories which arise from a tropical limit of conventional isotropic topological field theories. We analyze both the TBF theory and the tropical analogue of 2D topological Yang-Mills theory (TrYM) through a direct path integral calculation which probes a deformed analytic torsion and also through canonical quantization. The explicit construction of the Hilbert space of TrYM theory demonstrates that the TrYM theory provides an example of a solvable field theory where anisotropy properties and topological invariance can simultaneously hold. We show that the partition function has an asymptotic limit, which verifies that the dimension of the moduli space of tropicalized flat connection on a Riemann surface of genus $g>1$ is precisely given by $(g-1) \operatorname{rank}(\mathfrak{g})$. We show that the interpretation of this result is that the random matrix model associated to the U$(N)$ TrYM is in fact a novel random matrix theory whose integration space is still the same space of hermitian matrices (similar to a GUE) however the Dyson index matches that of a GOE consistent with the usual intuition that tropicalization reduces down complex objects to their real counterparts.
}
\begin{document}
\maketitle
\section{Introduction}
Modern theoretical physics has frequently shown that new research avenues begin by exploring frameworks that relax previously held assumptions that had been successful in the past. One such historically held assumption is that there is a fundamental spacetime symmetry underlying all quantum field theories known as relativistic invariance. However, it has been quite clear through the development of condensed matter physics \cite{altland2010condensed} and more recently in high energy physics that introducing an anisotropy between space and time vastly increases the theoretical landscape in model building.

This enlarged theoretical landscape has been particularly useful in the study of quantum gravity where the use of anisotropic scaling was employed to achieve a power-countable renormalizable theory of quantum gravity known as Hořava–Lifshitz gravity \cite{Horava:2008ih, Horava:2009uw}. By restricting the symmetry group to those preserving a preferred foliation of the underlying spacetime, this approach permits a richer set of spatial curvature terms while avoiding instabilities. This effectively sacrifices Lorentz symmetry at high energies, allowing one to construct models where gravitational interactions become well-behaved in the ultraviolet regime. 

Another intriguing manifestation of the anisotropic scaling emerges in the study of asymptotic symmetries \cite{Strominger:2017zoo} underpinning Carrollian physics. The asymptotic symmetries here arise in the ultra-relativistic limit where the speed of light is taken to zero; the resulting Carrollian framework effectively decouples temporal evolution from spatial dynamics, leading to a degenerate metric tensor and a symmetry group that departs significantly from the familiar Lorentz or Galilean cases. This anisotropic scaling is not merely a formal artifact; rather, it plays a central role in characterizing the asymptotic structure of spacetime, particularly at null boundaries. The enhanced symmetry groups capture the subtle interplay between geometric data at infinity and the local dynamics of the theory. 

By extending these ideas into the realm of string theory, one finds that non-relativistic formulations offer strikingly novel perspectives. Newton–Cartan strings \cite{ziqi}, for instance, reformulate the worldsheet dynamics by embedding them in a geometric framework defined by a degenerate metric that distinguishes between time and space, effectively relaxing Lorentz invariance. Similarly, Carrollian strings exhibit symmetry structures where temporal evolution decouples from spatial dynamics, leading to anisotropic scaling behaviors. It has been suggested that this can potentially provide novel insights into flat space holography \cite{Strominger:2013jfa, Stieberger:2024shv}. 

Alongside these developments, we have historically long-held the assumption in the 20th century that the only interesting high energy physics is present in perturbative scattering amplitudes. Topological field theories have since then emerged as a powerful paradigm where observables depend solely on global features rather than local geometrical details. This property enables exact computations and provides deep insights into nonperturbative phenomena ranging from topologically ordered phases in condensed matter systems to knot invariants in low-dimensional topology. Topological field theories have also emerged as essential tools in the exploration of relativistic quantum gravity toy models.  For example, three-dimensional Chern–Simons theory has been instrumental in capturing aspects of quantum geometry and black hole microstates, offering exactly solvable models where the interplay between topology and quantum mechanics is manifest. 

Given these two powerful paradigms, non-relativistic field theories and relativistic topological field theory, and motivated by the potential insights from studying anisotropic quantum gravity, it is natural to ask whether these approaches can be synthesized into a coherent framework that still employs the robust methods of topological field theory in a non-relativistic setting. In a recent work \cite{Albrychiewicz:2025yyg}, we have developed a simple prescription for constructing anisotropic Schwartz-type topological field theories in a similar spirit to the anisotropic Witten-type cohomological field theories \cite{ewtsm, ewcoho} that were constructed in \cite{Albrychiewicz:2023ngk}. 

The essential motivational principle behind the anisotropic scaling limits come from a well-known nonperturbative BPS configuration known as $(p,q)$ string networks \cite{sn,sni,snii,sniii,sniv}. These string networks arise as solitonic objects composed out of fundamental F-strings and D-1 strings which link up together in a piece-wise linear fashion with juncture conditions at their vertices identical to the piece-wise linear curves of tropical geometry. From an M-theory perspective, these string networks can lift up to a holomorphic curve in an 11-dimensional geometry \cite{krogh}. It was demonstrated that running this argument in reverse allows us to describe the underlying piece-wise linear curves through a singular, anisotropic limit of the holomorphic curves \cite{Ray:2008xq}. This is known as the tropical limit \cite{msintro, mikhalkinrau}. The essential observation is that one is no longer forced to work with the tropical curves themselves but can instead represent them using a foliated complex geometry equipped with nilpotent endomorphism of the tangent bundle known as Jordan structures. Consequently, writing down quantum field theories on these tropical curves becomes as simple as writing them down on foliated complex geometries with a few additional compatibility conditions on the fields so that they transform nicely under the foliation-preserving gauge symmetries that are introduced by the tropical limit.

The anisotropic topological field theory that we have focused on was the tropical analogue of the BF theory \cite{Schwarz:1978cn, Horowitz:1989ng, Blau:1989bq, Witten:1991we} in order to be able to quickly extend the prescription to more flexible topological field theories which can be built upon the results of BF theory. We have named this quantum field theory, the TBF theory \cite{Albrychiewicz:2025yyg} and in this paper, we present the analysis of the TBF partition function and the anisotropic holonomies that are associated with tropicalized flat connections as well as carefully analyze its connections to anisotropic conformal field theory.

In section \secref{sec:ClassGeomTBF}, we will give a brief recap on the construction of the underlying foliated geometry of TBF theory and obtain its Lagrangian through an anisotropic scaling limit of the original BF theory. This anisotropic scaling limit is implemented through what is known as the Maslov dequantization limit \cite{msintro, litvinov, viro, virohyper} which leads to tropical geometries that are represented as foliated complex geometries. We will take some time to discuss how to properly gauge fix the TBF theory through a generalization of the standard Hodge-star into a nilpotent operator that takes into account the underlying Jordan structure known as the Jordan-star $\star_J$. This Jordan star induces natural generalizations of the Hodge de-Rham Laplacian and allows us to make sense to the anisotropic tropical generalization of 2D topological Yang-Mills theory known as 2D TrYM theory.

In section \secref{sec:QuantTBF}, we will present the path integral quantization and canonical quantization of the TBF theory. In particular, we will analyze the path integral associated to the partition function on a cylinder through the Nicolai map \cite{Nicolai:1979nr, Cecotti:1981fu, Birmingham:1988bx, Blau:1989bq, Blau:1989dh} and explicitly show that it counts the moduli space of tropicalized flat connections to be $\operatorname{rank}(\mathfrak{g})$ on a sleeve geometry as was shown through a twisted cohomology argument in \cite{Albrychiewicz:2025yyg}. We perform the canonical quantization of the 2D TrYM theory on a general foliated Riemann surface $\Sigma_{g,b}$ of genus $g$ with $b$ boundaries by decomposing the problem into the simpler problem of analyzing the Hilbert space on foliated Riemann surfaces with a pair of pants topology. We find that the Hilbert space of states is effectively the same as in the usual 2D isotropic topological Yang-Mills \cite{Witten:1992xu, Cordes:1994fc} theory except for the fact that all holonomies can globally be put into the same Cartan subalgebra, which allows us to write down all holonomies as automatically commuting.  Using these results, we write the partition function in general for a Riemann surface of genus $g$ and show through a Poisson resummation that in the topological limit where the Yang-Mills coupling vanishes, the partition function correctly counts that the dimension of the moduli space of tropicalized flat connections is $(g-1)\operatorname{rank}(\mathfrak{g})$.

In section \secref{sec:Conclusions}, we conclude with a discussion on how the 2D TrYM admits an easily constructable matrix models with what is currently available in the literature \cite{Aganagic:2004js, Szabo:2010qv, Anninos:2020ccj}. We give some closing remarks that several of the standard questions that were asked for the standard isotropic topological Yang-Mills theory naturally have a corresponding question for its tropical analogue, TrYM theory. We also discuss how the main physical motivation behind the 2D TrYM theory lies in the conjecture that it might be a candidate for a quantum field theory that probes some of the anisotropic worldsheet physics that one would expect in the wedge region of non-equilibrium string perturbation theory.

\section{Classical Geometry of the TBF and Anisotropic Yang-Mills Theory}
\label{sec:ClassGeomTBF}

In this section, we review the essential calculations of how anisotropy can be encoded into path integrals through the Maslov dequantization limit of tropical geometry. For additional physical motivations and examples of why and how the Maslov dequantization limit is applied, please refer to \cite{Albrychiewicz:2023ngk, Albrychiewicz:2024tqe, Albrychiewicz:2025rkg, Albrychiewicz:2025yyg}. In particular, we will examine the construction of 2D TBF theory, its generalization to 2D anisotropic\textit{ tropological }Yang-Mills (TrYM) and discuss the construction of an analogous Hodge star which is suitable for path integrals and the canonical quantization of the theories.

\subsection{Essentials of Tropical Geometry and 2D TBF and 2D TrYM Theory}
As mentioned in the introduction, our motivations in using tropical geometry come from the observation that they are a physically natural way of describing anisotropic limits of field theories. Tropical geometry has previously made its appearance in physics in the study of BPS superstring configurations, Feynman integrals \cite{Arkani-Hamed:2022cqe}, homological mirror symmetry \cite{gross2011tropical} and in the study of superstring amplitudes \cite{Tourkine:2013rda, Eberhardt:2022zay}. Despite these, one can argue that there is still plenty potential applications of tropical geometry in physics. In short, tropical geometry is based on combinatorial piece-wise linear structures and by transforming classical algebraic geometric objects into polyhedral and combinatorial entities, it enables one to tackle sophisticated geometric questions with discrete tools.

The transformation is based on the semiring underlying tropical geometry i.e., the tropical semiring $\mathbb{T}$, where the arithmetic operations are defined as follows:
\[
a \oplus b = \max(a, b) \quad \text{and} \quad a \otimes b = a + b.
\]
Here, the tropical addition \(\oplus\) represents the maximum, and tropical multiplication \(\otimes\) is standard addition in $\mathbb{R}$. Notice that the operation \(\oplus\) is idempotent, which leads to vast simplifications when it can be applied. A tropical polynomial in one variable can be expressed as
\[
F(x) = \max\{a_0,\, a_1 + x,\, a_2 + 2x,\, \dots,\, a_n + nx\},
\]
where $n$ is the degree of the tropical polynomial. As an illustrative example of what these polynomials are, consider a tropical polynomial in two variables defined by
\[
F(x, y) = 0 \oplus x \oplus y = \max\{0,\, x,\, y\}.
\]
The tropical curve associated with this polynomial consists of the set of points \((x,y) \in \mathbb{R}^2\) where the maximum is achieved at least twice. In this case, the curve is determined by the conditions \(0 = x\), \(0 = y\), or \(x = y\), which yield three rays emanating from the origin. Consequently, from its definition, one can see that a tropical polynomial is essentially given by piecewise linear curves. 
 
This is in sharp contrast to how complex varieties are defined in terms of the vanishing of a smooth polynomial, for e.g., $F(x,y)=0$. This would make a physicist worry about using the tropical semiring for their usual operations of interest, such as gauge fixing, which is for example given by the condition $F(x,y)=0$. Another conceptual obstacle concerns the encoding of the tropical semiring into the corresponding path integrals: must one evaluate these path integrals in terms of min/max operations?  As argued in \cite{Albrychiewicz:2023ngk}, it turns out that this is not the case; we are still able to work with our conventional path integrals defined over fields such as $\mathbb{C}$. 
 
From here onward, we will present the tropicalization prescription for the case of quantum field theories that are defined over a Riemann surface $\hat\Sigma$ with a complex structure $\hat{J}$ prior to taking the tropical limit. For the sake of completeness, we also equip the Riemann surface with a metric tensor $\hat{g}$ on the tangent bundle $T\hat{\Sigma}$ and a cometric tensor $\hat{h}$ on the cotangent bundle $T^*\Sigma$ which is naturally identified as the inverse $\hat{g}^{-1}=\hat{h}$. The reason we make a distinction between the metric tensor and the cometric is due to the fact that they will no longer be inverses after the tropical limit. Once the tropical limit is taken, we will drop the corresponding hats in the notation.

We begin by employing local complex coordinates $(z,\bar{z})$ on $\hat{\Sigma}$. The tropical limit of the complex geometry can be implemented by taking the $\hbar\rightarrow 0$ limit (the Maslov dequantization) of the subtropical deformation of the field of the complex numbers
\begin{align}
    S_\hbar(z)=\begin{cases}
    |z|^{1/\hbar}\frac{z}{|z|}, \quad \, \text{if }z\neq 0, \\
    0, \qquad \qquad \text{if }z=0.
    \end{cases}
\end{align}
For complex coordinates, parametrized using polar variables $(r,\theta)$ the subtropical deformation gives
\begin{equation}
\label{eqn:Maslov}
z=e^{\frac{r}{\hbar}+i \theta}, \quad \bar{z}=e^{\frac{r}{\hbar}-i \theta},
\end{equation}
and in the tropical limit, the complex structure $\hat{J}$ degenerates into a nilpotent endomorphism of the tangent bundle $J:T\Sigma \rightarrow T\Sigma$ known as Jordan structure whose action in the adapted coordinates $(r,\theta)$ defined by the Maslov dequantization is
\begin{equation}
J\left(\partial_r\right)=\partial_\theta, \quad J\left(\partial_\theta\right)=0.
\end{equation}
One can construct the corresponding dual map $J^*: T^*\Sigma\rightarrow T^*\Sigma$ in adapted coordinates via the action
\begin{equation}
J^*(d \theta)=d r, \quad J^*(d r)=0.
\end{equation}

\begin{figure}[ht]
    \centering
    \includegraphics[width=0.8\linewidth]{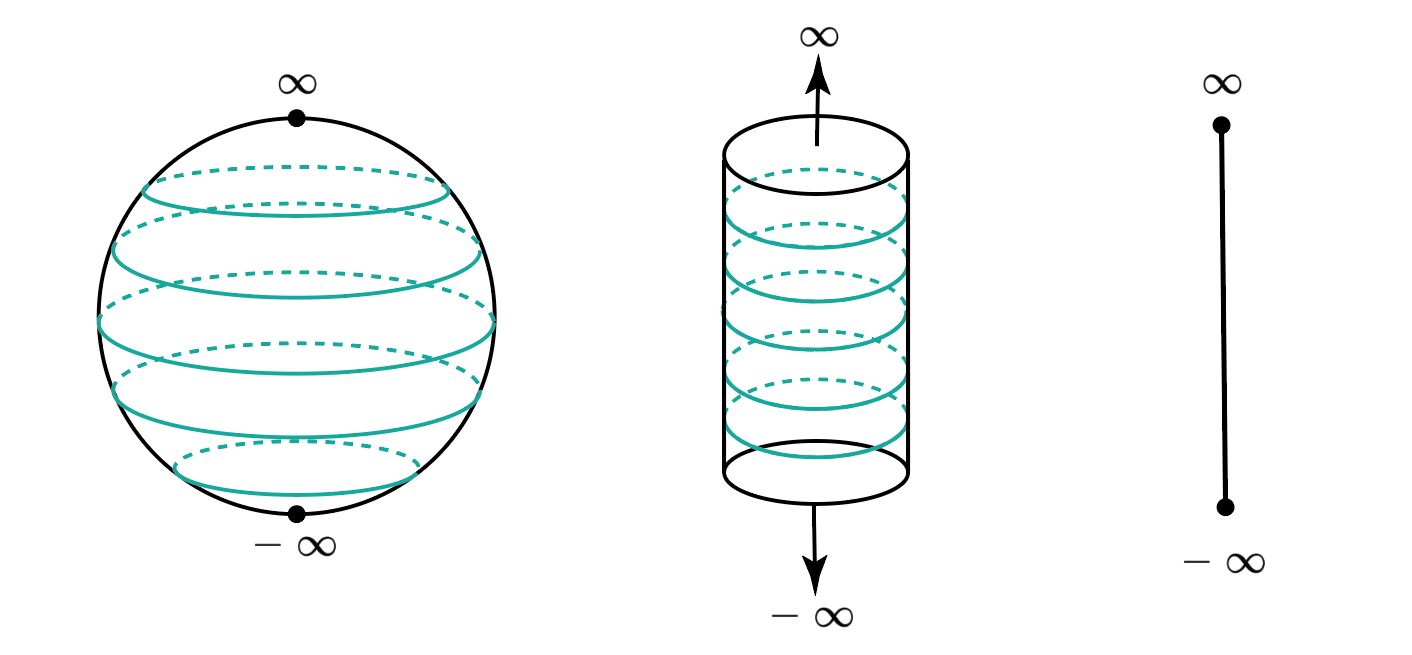}
    \caption{The three perspectives of the underlying tropical geometry. The leftmost picture shows tropical projective space as a foliated $\mathbb{CP}^1$, the middle picture shows its equivalency to an infinite foliated cylinder known as the sleeve and the rightmost picture would be the quotient space that on would obtain from collapsing the space of leaves into the underlying tropical graph. All three perspectives are equivalent for the purposes of quantum field theory, we call the sleeve the covering space perspective and the one-dimensional tropical graph, the quotient space perspective.}
    \label{fig:sleeve1}
\end{figure}

Due to the fact that this limit is taken in terms of local coordinates, we are effectively working on a cylinder known as a sleeve, see \figref{fig:sleeve1}, since $\theta$ is a periodic coordinate. Later, we will see that we can glue several sleeves together in order to form a more topologically complicated surfaces. Unlike a standard complex structure which provides a splitting of the complexified tangent space into holomorphic and antiholomorphic components, the Jordan structure instead gives us a two-step filtration on the tangent space
\begin{equation}
0 \subset \operatorname{im}(J) = \operatorname{ker}(J) \subset T \Sigma.
\end{equation}
We want to emphasize that in taking the tropical limit, we do not change the differential structure that defines the underlying manifold; any objects constructed purely from the differential structure, such as exterior derivatives, differential forms and associated bundles, remain the same. There is an important caveat to this statement; the Jordan structure defines a distribution on $\Sigma$ which is automatically integrable in the two-dimensional case, and hence, defines a foliation on $\Sigma$.  If one asks what sort of diffeomorphisms preserve the structure of the Jordan structure or interchangeably, the foliation, then one finds foliation-preserving diffeomorphisms
\begin{equation}
\begin{aligned}
& \widetilde{r}=\widetilde{r}(r), \\
& \widetilde{\theta}=\widetilde{\theta}_0(r)+\theta \partial_r \widetilde{r}(r).
\end{aligned}
\end{equation}
By looking at the Lie algebra associated to this symmetry group, we obtain
\begin{equation}
\begin{aligned}
\delta r & =f(r), \\
\delta \theta & =F(r)+\theta \partial_r f(r),
\end{aligned}
\end{equation}
thus, the local symmetries of the Jordan structure are generated by the infinite-dimensional Lie algebra whose elements are parametrized by two real, arbitrary, projectable, and differentiable functions $f(r)$ and $F(r)$ on the foliation. As a result, tropical geometries can be recovered in the tropical limit i.e., $\hbar\rightarrow 0$ as a foliated complex geometry. We take advantage of this perspective and construct anisotropic quantum field theories that live on geometries which are described by tropical geometry in the sense that they naturally live on foliated complex geometries or more generally filtered manifolds \cite{Albrychiewicz:2025afk}.

Recall that we can equip our initial Riemann surface $\hat{\Sigma}$ with a nondegenerate metric tensor $\hat{g}$ and a nondegenerate cometric $\hat{h}$. In the tropical limit, both of these objects transform into degenerate bilinear forms so that in the adapted coordinates they have the following matrix representations
\begin{equation}
g=\left(\begin{array}{ll}
1 & 0 \\
0 & 0
\end{array}\right), \quad h=\left(\begin{array}{ll}
0 & 0 \\
0 & 1
\end{array}\right).
\end{equation}
As a result, we no longer have the standard musical isomorphisms between the tangent and cotangent bundle. Instead, we have a mutual invisibility condition between the tropical metric and the tropical cometric $hg=gh=0$.  For the case of complex projective space $\mathbb{CP}^1$, we can intuitively see how the tropical limit is implemented through the $\hbar\rightarrow 0$ limit of the Maslov dequantization.  The resulting geometry is tropical projective space $\mathbb{TP}^1$, see also \figref{fig:sleeve1}.  

Although, in principle, we will formulate TBF and TrYM theory on sleeves, we will find it useful to extend these sleeves to infinity for certain analyses and instead formulate the theory on $\mathbb{TP}^1$. If we want to glue several of these sleeves together in order to obtain a foliated Riemann surface of genus $g$, then we will emphasize the sleeve perspective. The 2D TBF theory \cite{Albrychiewicz:2025yyg} is then constructed by taking a tropical limit of the 2D BF theory whose Lagrangian action is
\begin{equation}
S_{B F}[B, A]=\int_{\hat{\Sigma}} \operatorname{tr}(B \wedge F),
\end{equation}
where $F=d A+A \wedge A$ is the curvature two form on a 2-dimensional manifold $\hat\Sigma$ and the $\operatorname{tr}$ is the trace taken in the adjoint representation of the Lie algebra $\mathfrak{g}$ of the gauge group $G$ and $B$ is an adjoint-valued 0-form. In order to take the limit properly, one must supplement the Maslov dequantization procedure with a deformation of the fields. In the adapted coordinates, the prescription for deforming the components of fields becomes
\begin{equation}
\begin{aligned}
A_r & \rightarrow A_r, \\
A_\theta & \rightarrow \hbar A_\theta, \\
B & \rightarrow \frac{1}{\hbar} B,
\end{aligned}
\end{equation}
the TBF action after a Hubbard-Stratonovich transformation and sending $\hbar\rightarrow0$ is then
\begin{equation}
\label{eqn:TropicalTBFAct}
S_{T B F}=\int_{\Sigma} d r \wedge d \theta \operatorname{tr}\left[T\left(-\partial_\theta A_r\right)+B\left(\partial_r A_\theta+\left[A_r, A_\theta\right]\right)\right].
\end{equation}
Here, the Hubbard-Stratonovich transformation introduces a new adjoint-valued scalar field $T$. 

The flatness condition of the BF theory gets deformed into a transverse flatness condition given by
\begin{equation}
f_{r\theta}=\partial_r A_\theta+\left[A_r, A_\theta\right]=0,
\end{equation}
and the equation of motion of $T$ imposes a projectability condition on the radial component of the tropicalized connection $A$ which is
\begin{equation}
    \partial_\theta A_r=0.
\end{equation}
The Euler-Lagrange equations for the components $A_r$ and $A_\theta$ give
\begin{equation}
\begin{aligned}
& \partial_\theta T+\left[A_\theta, B\right]=0, \\
& \partial_r B+\left[A_r,B \right]=0.
\end{aligned}
\end{equation}
The TBF action is invariant under the following foliation-preserving gauge symmetries
\begin{equation}
\begin{gathered}
\label{eqn:GaugeSym}
\delta_\eta A_r=\partial_r \eta+\left[A_r, \eta\right], \quad \delta_\eta A_\theta=\left[A_\theta, \eta\right], \\
\delta_\eta T=[T, \eta], \quad \delta_\eta B=[B, \eta],
\end{gathered}
\end{equation}
where $\eta$ is a $\theta$-independent, adjoint-valued scalar.  Unlike 2D BF theory, we have an additional anisotropic topological shift symmetry parametrized by a projectable Lie-algebra valued scalar $\lambda(r)$ given by 
\begin{equation}
\begin{gathered}
\label{eqn:AddSym}
\delta_\lambda A_r=0, \quad \delta_\lambda A_\theta=0, \\
\delta_\lambda T^a=\lambda^a(r), \quad \delta_\lambda B=0,
\end{gathered}
\end{equation}
when the $\theta$-direction is periodic. In order to extend this to tropical topological Yang-Mills theory, we will need to construct the tropical analogue of the Hodge star. 

\subsection{The Jordan Star and Gauge Fixing Conditions}
Before analyzing any more sophisticated observables, it is instructive to know how the direct path integral quantization works. The partition function of the TBF theory and the partition function of the TrYM theory at zero coupling will coincide. As was shown in \cite{Albrychiewicz:2025yyg}, the partition function of 2D TBF theory can be shown to localize onto the moduli space of tropicalized flat connections identified under foliation-preserving gauge transformations. In order to perform the path integral of the partition function, we want to gauge fix the foliation-preserving gauge transformations. 

We implement this through standard cohomological BRST methods. We begin by defining a nilpotent graded differential operator on the space of fields known as the BRST differential $Q$ whose action is
\begin{equation}
\begin{aligned}
&Q A_r=\partial_r c+\left[A_r, c\right] ,\\
&Q A_\theta=\left[A_\theta, c\right] , \\
&Q T = [T,c] , \\
&Q B = [B,c] , 
\end{aligned}
\end{equation}
Here, the ghost field $c$ is a projectable, adjoint-valued, Grassmann odd field $c(r)$. The BRST differentials acts on the ghost fields such that the operator is ensured to be nilpotent.
\begin{equation}
Q c=-\frac{1}{2}[c, c].
\end{equation}
In order to design a proper gauge fixing fermion, we want to take a look at the gauge fixing condition of the BF theory which is the divergence-less condition $d_\Phi^{\dagger} A=0$, where $\Phi$ is some 1-form section from an auxiliary vector bundle.

In order to formulate the tropical analogue of the divergenceless condition, we need to introduce an appropriate notion of adjoint derivative which is based on two ingredients: an inner product on the space of differential forms compatible with the foliation-preserving diffeomorphisms and a generalized notion of Hodge star.  Recall that in taking the tropical limit, the metric tensor $g$ becomes degenerate and the usual methods which rely on a non-degenerate metric tensor are lost. For the case of a 2-real dimensional manifold $\Sigma$ with adapted coordinates $(r,\theta)$, the line element takes the form
\begin{equation}
d s^2=d r^2.
\end{equation}
One of the consequences of this degeneracy is that the Riemannian volume form $\omega$ vanishes. As a result, one is not able to write down a Hodge star operator that is induced by the metric tensor which is usually defined through the Riemannian volume form as 
\begin{equation}
\alpha \wedge \star \beta=\langle\alpha, \beta\rangle \omega,
\end{equation}
Here, $\alpha$ and $\beta$ are differential forms of the same deRham degree, and the inner product is the canonical inner product on the exterior algebra induced by a non-degenerate metric tensor. One last consequence of this is that without a Hodge dual, one is no longer able to leverage conventional Hodge theory in order to construct Hodge-deRham Laplacian whose functional determinant is usually of interest, nor can one impose the covariant gauge.

Despite us not having a non-degenerate metric tensor, it turns out that we can still construct a useful analogue of the Hodge star from the Jordan structure. Recall that the dual action of the Jordan structure is given by $J^*: T^*\Sigma\rightarrow T^*\Sigma$ in adapted coordinates via the action
\begin{equation}
J^*(d \theta)=d r, \quad J^*(d r)=0.
\end{equation}
From this, we define the \textit{Jordan star} operator  $\star_J=-J^*$ on 1-forms. We can evaluate this to be
\begin{equation}
\star_J d r=0, \quad \star_J d \theta=-d r.
\end{equation}
We extend this linearly for 1-forms $\alpha, \beta$, and find that this evaluates to
\begin{equation}
\alpha \wedge\star_J \beta=\alpha_\theta \beta_\theta d r \wedge d \theta,
\end{equation}
which induces a global foliation-preserving diffeomorphism invariant inner product on the differential 1-forms 
\begin{equation}
\langle\alpha \mid \beta\rangle_J=\int_{\Sigma} \alpha \wedge *_J \beta=\int_{\Sigma} \alpha_i \beta_j J_k^j d x^i \wedge d x^k= \int_{\Sigma} d r \wedge d \theta \hspace{0.1cm}\alpha_\theta \beta_\theta.
\end{equation}
This reflects the structure of the underlying tropical co-metric $h=d\theta \otimes d\theta$. 

The Jordan star operator does not satisfy the usual involution property characteristic of Hodge stars but instead satisfies the nilpotency condition
\begin{align}
    \star_J\star_J\alpha=0,
\end{align}
for any 1-form $\alpha$. The action of the Jordan star on 0-forms and 2-forms is
\begin{equation}
\begin{aligned}
\label{eqn:JordanStar02Forms}
& \star_J 1=dr \wedge d \theta ,\\
& \star_J (dr\wedge d \theta )=1.
\end{aligned}
\end{equation}
For the purposes of the anisotropic topological field theories, the top-form induced by the Jordan structure is sufficient to write down actions. The action of the Jordan star on 0-forms and 2-forms might not be so surprising since the standard Hodge star only depends on the integration measure instead of the metric tensor for the 2D topological theories that we are interested in. It is worth noting that there have been previous attempts at constructing a generalization of the Hodge star that is compatible with the data of a Carrollian manifold \cite{Fecko:2022shq}, the Jordan star is a third generalization that matches the same nilpotency property that Galilean and Carrollian Hodge star operators also satisfy.

With a respect to the natural foliation-preserving inner products on the space of 0-forms and 1-forms, one can compute the Jordan adjoint of the exterior derivative on 1-forms via
\begin{equation}
\left\langle -d^{\dagger_J} \alpha \mid f\right\rangle_J=\langle\alpha \mid d f\rangle_J,
\end{equation}
where we have inserted a conventional minus sign into the definition of the adjoint. A direct calculations shows that
\begin{equation}
d^{\dagger_J}=\star_J d \star_J,
\end{equation}
which can be evaluated on 1-forms $\alpha =\alpha_r dr +\alpha_\theta d\theta$ to yield the divergence operator along the leaves of the foliation
\begin{equation}
d^{\dagger_J} \alpha=\star_J d \star_J \alpha=\partial_\theta\alpha_\theta.
\end{equation}
The Jordan adjoint exterior derivative vanishes on 0-forms and consequently just like the exterior derivative, the Jordan adjoint of the exterior derivative is also nilpotent. We define the \textit{Jordan Laplacian} as
\begin{equation}
\label{eqn:JordanLaplace}
\Delta^J=d d^{\dagger_J}+d^{\dagger_J} d.
\end{equation}
On 0-forms, this can be evaluated to give
\begin{equation}
\Delta^J f=\partial_\theta^2 f.
\end{equation}
Whereas, on 1-forms, the Jordan Laplacian becomes
\begin{equation}
\Delta^J \alpha=\partial_\theta^2 \alpha_r d r+\partial_\theta^2 \alpha_\theta d \theta.
\end{equation}
In order to be able to apply this to TBF theory, we need to twist exterior derivatives so that it acts on Lie-algebra valued 1-forms $\alpha$ as
\begin{equation}
d_{A }\alpha=d \alpha+[A, \alpha],
\end{equation}
the Jordan adjoint of the twisted exterior derivative can then be evaluated to be
\begin{equation}
\label{eqn:NewAdjOneForms}
d_A^{\dagger_J} \alpha=\partial_\theta \alpha_\theta+\left[A_{\theta,} \alpha_\theta\right].
\end{equation}
Consequently, the tropical analogue of covariant gauge is  $d_A^{\dagger_J} \alpha=\partial_\theta \alpha_\theta+\left[A_{\theta,} \alpha_\theta\right]=0.$ Another useful result of the Jordan star is that we can now also directly write down the TrYM action as
\begin{equation}
S_{T r Y M}=\int_{\Sigma} d r \wedge d \theta \operatorname{Tr}\left[\frac{1}{2} T\left(-\partial_\theta A_r\right)+\frac{1}{2} B\left(\partial_r A_\theta+\left[A_r, A_\theta\right]\right)\right]-\frac{g^2}{4} \int_{\Sigma} \operatorname{Tr}\left(B \wedge \star_J B\right) .
\end{equation}

\section{Quantization of the TBF and Anisotropic Yang-Mills Theory}
\label{sec:QuantTBF}
In this section, we perform the explicit path integral quantization of partition function of the 2D TBF theory on a sleeve and show that it correctly picks out $\operatorname{rank}{\mathfrak{g}}$ degrees of freedom. Then, we extend the 2D TBF theory to the tropical analogue of 2D topological Yang-Mills theory and show how to explicitly canonically quantize the resulting theory with the additional $T$ field. We argue that this analysis extends to a general foliated Riemann surface $\Sigma_{g,b}$ of genus $g$ and $b$ boundaries. From this, we are able to extract the dimension of the moduli space of tropicalized flat connections in the topological limit where the Yang-Mills coupling $e$ vanishes.

\subsection{The Partition Function and Deformed Analytic Torsion}
\label{subsec:torsion}

The partition function of tropical BF theory can be computed analogously to the relativistic case. In particular, we argue that the partition function is exact; namely, it depends solely on deformed flat connections and receives no contributions from ghost terms. For this purpose, we use the Jordan star and the corresponding Jordan adjoint of any relevant operator
\eqref{eqn:NewAdjOneForms} (introduced in \secref{sec:ClassGeomTBF}) to define a ``tropical” analogue of what is conventionally referred to as a covariant gauge \cite{Witten:1989sx}.

We start by expanding the action \eqref{eqn:TropicalTBFAct} around the classical solutions
\begin{align}
    A_r&=A_r^c+A_r^q, \\ \nonumber
    A_\theta&=A_\theta^c+A_\theta^q, \\ \nonumber
    B&=B^c+B^q, \\ \nonumber
    T&=T^c+T^q,
\end{align}
which satisfy the equations
\begin{align}
    \partial_r A^c_\theta&+[A_r^c,A_\theta^c]=0, \\ \nonumber
    \partial_\theta A_r^c&=0, \\ \nonumber 
    \partial_\theta T^c&+[A_\theta^c, B^c]=0, \\ \nonumber
    \partial_r B^c&-[B^c,A_r^c]=0. 
\end{align}
We obtain 
\begin{align}
\label{eqn:QCactionCoords}
    S_0&=\int_\Sigma\, dr \wedge d\theta\left(-T^q\partial_\theta A_r^q+ B^q(\partial_r A_\theta^q+[A_r^q,A_\theta^c]+ [A_r^q,A_\theta^q])\right),
\end{align}
where $B^c$ and $T^c$ zero modes are eliminated by using delta constraints that arise once one integrates the fluctuation fields $B^q$ and $T^q$. We can rewrite the action
 \eqref{eqn:QCactionCoords} in terms of an exterior derivative which twisted by the radial projection of the tropical connection $A_rdr$. The action of the twisted exterior derivative is given by
\begin{align}
 d_{(A_r dr)}(A_\theta d\theta)=\left(\partial_r A_\theta^q+[A_r^q,A_\theta^c]+[A_r^q,A_\theta^q]\right)dr\wedge d\theta,
\end{align}
and
\begin{align}
    d_{(A_r dr)}(A_r dr)=-\partial_\theta A_r^q dr\wedge d\theta,
\end{align}
where we use the gauge fixing condition for the classical solution $A_r^c=0$ as discussed before.  The gauge fixing condition for the fluctuation fields is then given by
\begin{align}
    d_{A^c}^{\dagger_J} A^q=\partial_\theta A_\theta^q+[A_\theta^c,A_\theta^q].
\end{align}
In order to construct the Lagrangian action for the ghost sector, we note that under the foliation preserving gauge transformations, we have
\begin{align}
    \delta_\eta(d_{A^c}^{\dagger_J} A^q)=\partial_\theta[A_\theta,\eta]+[A_\theta^c,[A_\theta,\eta]],
\end{align}
where $\eta$ is a $\theta$ independent symmetry generator. As a result, the kinetic term for the ghost $c$ and antighost fields $\bar{c}$ is given by
\begin{align}
S_g= \int_\Sigma  dr \wedge d\theta \hspace{0.1cm}   \bar{c}(r)d_{A_\theta^cd\theta}^{\dagger_J} d_{(A_\theta d\theta)}c(r), 
\end{align}
here the ghost fields are projectable fields in the sense that they are also $\theta$ independent due to the fact that underlying gauge symmetry is a projectable one.  The angular dependency of the Lagrangian density comes in from the $A_\theta$ fields. In summary, the gauge fixed action which we use to compute the 2D TBF partition function becomes
\begin{align}
    S&=\int_{\Sigma} dr\wedge d\theta \hspace{0.2cm} T^qd_{(A_r dr)}(A_r dr)+B^qd_{(A_r dr)}(A_\theta d\theta)+\lambda d_{A^c}^{\dagger_J} A^q+\bar{c}d_{A^c}^{\dagger_J} d_{(A_\theta d\theta)}c.
\end{align}
As in the relativistic case, we argue that this partition function localizes on the moduli space of flat connections. We demonstrate this by introducing Nicolai map \cite{Nicolai:1979nr, Cecotti:1981fu, Birmingham:1988bx, Blau:1989bq, Blau:1989dh}
\begin{align}
\label{eqn:NicolaiFields}
    \xi(A)=d_{(A_rdr)}(A^c+A^q), \quad \eta(A)=d^{\dagger_J}_{A_c}A^q,
\end{align}
which simplifies the bosonic part of the action and leaves the ghost term unchanged. Then, we obtain
\begin{align}
    S&=\int_\Sigma dr \wedge d\theta \hspace{0.2cm} \CB \xi+\lambda \eta+\bar{c}d_{A^c}^{\dagger_J} d_{(A_\theta d\theta)}c,
\end{align}
where $\CB=(T^q,B^q)$. The change of variables \eqref{eqn:NicolaiFields} introduces a Jacobian into the path integral which has the form
\begin{align}
    \det\left(\frac{\delta(\xi, \eta)}{\delta A}\right)^{-1}=\text{det}(d_{A^c}^{\dagger_J} d_{(A_r dr)})^{-1}.
\end{align}
The functional integration over $\CB$ and $\lambda$ coupled with the fact that classical solutions satisfy the transverse flatness conditions, forces their fluctuations to vanish. Whereas, the functional integration of the ghost fields leads to the ratio of determinants
\begin{align}
\label{eqn:RatioOfDets}
    \frac{\text{det}'(d_{A^c}^{\dagger_J} d_{A^c})}{\text{det}'(d_{A^c}^{\dagger_J} d_{A^c})},
\end{align}
which evaluates to the identity upon removing covariantly constant zero modes from determinants. The leftover finite-dimensional integral is over the moduli space of tropicalized flat connections since there is a one-to-one correspondence between the fixed points of the Nicolai map and points in the moduli space of tropicalized flat connections. In the case when the corresponding gauge group is a compact Lie group, then this is a finite dimensional integral over the Cartan subalgebra. We can parametrize the Cartan subalgebra with coordinates $a_i$ and expand the corresponding Lie-algebra valued 1-forms as
\begin{align}
    A^c=\sum_{i=1}^ra_iH^id\theta.
\end{align}
In order to explicitly perform the integral, we rewrite the measure on the moduli space tropicalized flat connections in terms of the standard Lebesgue measure on the Cartan subalgebra $\mathfrak{h}$
\begin{align}
    d\mu_{\CM}(a)=\frac{1}{|W|}\prod_{i=1}^r\frac{da^i}{2\pi},
\end{align}
where we quotient by the order of the Weyl group $W$ to account for large gauge transformations. Hence, our result
\begin{align}
    Z_{TBF}=\int_{\CM}d\mu_{\CM}(a)=\text{Vol}(\CM(M,G)),
\end{align}
is in agreement with the observation made in \cite{Albrychiewicz:2025yyg} that the partition function of TBF theory localizes on the moduli space of tropicalized flat connections. 

We now focus on the theory defined on $M=\mathbb{TP}^1$ with gauge group $G$. The real dimension of the moduli space can be directly computed:
\begin{align}
    \dim(\CM(\mathbb{TP}^1,G))=\text{rank}(\mathfrak{g}). 
\end{align}
Using the Jordan adjoint, we can connect the ratio of determinants \eqref{eqn:RatioOfDets} that appeared from the Jacobi map with a deformed analytic torsion \cite{RAY1971145, Schwarz:1978cn}. 
\begin{align}
    \tau(M)=\prod_{p=0}^d\left(\text{det}'\Delta_p^J\right)^{-(-1)^p\frac{p}{2}},
\end{align}
where we introduced the Jordan Laplacian on the space of p-forms
\begin{align}
    \Delta_p^J=d^{\dagger_J}d+d d^{\dagger_J},
\end{align}
and assumed that zero modes were removed from the functional determinant. For our particular 2D case, we get
\begin{align}
    \tau(M)= \frac{(\det' \Delta_1^J)^{1/2}}{\det' \Delta_2^J},
\end{align}
and we argue below, based on the spectrum of these two operators, that
\begin{align}
\label{eqn:EquivOpTor}
    \det \Delta^J_1=\det (\Delta_2^J)^2, 
\end{align}
which suggests that the torsion vanishes and its equivalency to \eqref{eqn:RatioOfDets}. 

Consider a non-harmonic 1-form $\eta$ with respect to the Jordan Laplacian, then one can show through direct calculation that it admits a decomposition of the form 
\begin{align}
    \eta=df+d^{\dagger_J}\omega,
\end{align}
for a 0-form $f$ and 2-form $\omega$. Now suppose that f is an eigenfunction of the Jordan Laplacian on 0-forms and denote its eigenvalue by $\lambda$
\begin{align}
    \Delta^J_0f=\lambda f,
\end{align}
then 
\begin{align}
    \Delta^J_2(\star_J f)=\star_J\Delta_0^Jf, 
\end{align}
which implies that the spectrum of eigenvalues of these two operators are the same. Combined with observations that
\begin{align}
    \Delta_1^J\eta=d\Delta_0 f+d^\dagger_J\Delta_2\omega,
\end{align}
we confirm \eqref{eqn:EquivOpTor}.
\subsection{The TrYM Hilbert Space via Canonical Quantization}
Similarly to 2D isotropic topological Yang-Mills, the most interesting class of observables beyond the partition function will be the Wilson loop insertion. Unlike the partition function which admits a straightforward path integral computation, we begin with the canonical quantization of the tropical topological Yang-Mills theory (TrYM) along the lines of \cite{Cordes:1994fc}.  We will see that for the standard 2D isotropic topological Yang-Mils, the tropical analogue will only see ``half" of the number of degrees of freedom on higher genus Riemann surfaces, and even for the non-abelian gauge group the holonomies of connections will commute.

To begin, we describe the tropical topological Yang-Mills theory in the second order formalism as
\begin{equation}
S_{T r Y M}=\int_{\Sigma} d r \wedge d \theta \operatorname{Tr}\left[\frac{1}{2} T\left(-\partial_\theta A_r\right)+\frac{1}{2 e^2}\left(\partial_r A_\theta+\left[A_r, A_\theta\right]\right)^2\right] ,
\end{equation}
the equations of motion of this action are

\begin{align}
\label{eqn:Const1}
& \partial_\theta A_r=0, \\ \label{eqn:Const2}
& \frac{1}{2} \partial_\theta T+\frac{1}{e^2}\left[A_\theta,\left(\partial_r A_\theta+\left[A_r, A_\theta\right]\right)\right]=0, \\ 
& \partial_r\left(\partial_r A_\theta+\left[A_r, A_\theta\right]\right)+\left[A_r, \partial_r A_\theta+\left[A_r, A_\theta\right]\right]=0 .
\end{align}    
We can identify the first two equations as constraint equations that we will impose on the gauge invariant wave functionals $\Psi[A_r,A_\theta,T]$ that describe the Hilbert space of this theory. As was stated in \cite{Albrychiewicz:2025yyg}, it is possible to gauge fix \eqref{eqn:GaugeSym} such that $A_r=0$ and $A_\theta= A^l_\theta(r,\theta)H_l$ lies in the Cartan subalgebra, where $H_l$ are generators of the Cartan subalgebra and $l$ is an index that runs from $\{{1,...,\operatorname{rank}(\mathfrak{g})\}}$. The gauge fixing condition $A_r=0$ eliminates the first constraint \eqref{eqn:Const1}. The adjoint-valued conjugate momenta is
\begin{equation}
e^2\pi_{A_{\theta}}^a=\partial_r A_\theta^a+\left[A_r, A_\theta\right]^a=\partial_rA_\theta^a,
\end{equation}
after gauge fixing $A_r=0$, the quantization constraint \eqref{eqn:Const2} appears as
\begin{equation}
\begin{gathered}
\left(-\frac{e^2}{2}  T^a(\theta)\partial_\theta+f_{b c}^a A_\theta^b(\theta) \frac{\delta}{\delta A_\theta^c(\theta)}\right) \Psi[A_\theta,T]=0,
\end{gathered}
\end{equation}
where our convention for the structure constants of the Lie algebra is $[t_a,t_b]=f^c_{ab}t_c$.  Define the differential operators
\begin{equation}
L^a=f_{b c}^a A_\theta^b \frac{\delta}{\delta A_\theta^b}\,,
\end{equation}
their commutator is well known and can be explicitly calculated to give a representation of the Lie-algebra $\left[L^a, L^b\right]=f_c^{a b} L^c$ through the Jacobi identity $f^{a b}{ }_c \equiv f^a{ }_{c d} f^b{ }_{d e}-f^b{ }_{c d} f^a{ }_{d e}$. Using this, we can rewrite this quantization constraint as 
\begin{equation}
L^a \Psi=\frac{e^2}{2} \partial_\theta T^a \Psi.
\end{equation}
Letting the commutator $[L_a,L_b]$ act on the wave functionals and using the commutation relations gives the constraint
\begin{equation}
\partial_\theta T^a=0.
\end{equation}
Therefore, on a spatial slice, we will only get constant modes $\tau$ in the Cartan sub-algebra for the $T$ field. 

Nontrivial wave functionals are those that satisfy
\begin{equation}
\begin{gathered}
\left(-\frac{e^2}{2}  \tau\partial_\theta+f_{b c}^a A_\theta^b(r,\theta) \frac{\delta}{\delta A_\theta^c(r,\theta)}\right) \Psi[A_\theta,T]=0.
\end{gathered}
\end{equation}
We recall that, away from the spatial slice that defines the canonical quantization, we have an additional anisotropic topological symmetry \eqref{eqn:AddSym}, which acts on $\tau^a(r)$ as
\begin{equation}
\delta_\lambda \tau^a=\lambda^a(r).
\end{equation}
As a result, we can completely gauge away the $T$ field on our physical wave functionals.  The Hilbert space is then constructed from wave functionals that satisfy
\begin{equation}
\begin{gathered}
\left(f_{b c}^a A_\theta^b(\theta) \frac{\delta}{\delta A_\theta^c(\theta)}\right) \Psi[A_\theta]=0,
\end{gathered}
\end{equation}
which states that physical states must be gauge invariant under the same transformation laws that $A_\theta$ obeys.  In the same vein as 2D topological Yang-mills, the natural class of wave functionals are then traces of holonomies
\begin{equation}
\Psi= \operatorname{Tr} e^{\int_0^{2\pi} d \theta A_\theta( \theta)}.
\end{equation}
The main difference of this wave functional with regards to 2D isotropic topological Yang-Mills theory is that these wave functionals are composed out of tropical connections that always lie in the Cartan subalgebra meaning that holonomies will generally commute. The wave functional can therefore be written as characters $\chi_R$ of exponentials of Cartan generators in an irreducible representation $R$ of the gauge group $G$ of foliation-preserving gauge transformations:
\begin{equation}
\label{eqn:Wavefncs}
\Psi[A_\theta]=
 \hspace{0.1cm}\chi_R\left(e^{2 \pi i a^l H_l}\right)= \chi_R\left(a^1, \ldots, a^{\operatorname{rank}(\mathfrak{g})}\right).
\end{equation}
From this result, we can see that the character component of the total wavefunction only depends on the boundary holonomy $g=e^{2 \pi i a^l H_l}\in G$ which is valued in a finite-dimensional unitary representation $R$ of $G$. Consequently, they define elements of the vector space $\Omega^0(G, R)^G$ of $G$-equivariant $R$-valued functions or equivalently $G$-invariant smooth functions on the gauge group of foliation preserving gauge transformations. We explicitly have $\operatorname{rank}\mathfrak{g}$ degrees of freedom since the tropical component $A_\theta$ and their corresponding holonomies all commute. As usual, we will restrict to $L^2$ integrable wavefunctions and are leftover with $L^2(G, R)^G$.  The corresponding Hilbert space $\mathcal{H}$ of 2D TrYM is then
\begin{equation}
\mathcal{H}= L^2\left(\Omega^0(G, R)^G\right).
\end{equation}

In order to fully characterize this Hilbert space $\mathcal{{H}}$, we have to construct an G-invariant inner product that respects the foliation and for which Hilbert spaces are square-integrable with respect to. The gauge group of foliation-preserving gauge transformations satisfy the additional projectability condition $\partial_\theta g=0$, which restricts it to only elements in the Lie-group which lie in the kernel of the differential 1-form $\omega=\partial_\theta g$. The kernel of this differential 1-form is a one-dimensional distribution, which is automatically integrable on any Lie-group and thus correspondingly also defines a foliation on the gauge group. Fortunately, the construction for the Haar measure does not change under this additional data and our inner products for $\mathcal{H}=L^2\left(\Omega^0(G, R)^G\right)$ can still be equipped with the standard Haar measure.  Generically, the class of test functions that we would have to consider would be the natural projectable ones; however, we have seen above that wavefunctions \eqref{eqn:Wavefncs} are written in terms of zero-modes and therefore the group integration is not explicitly affected by the projectability conditions.

Since $R$ is a unitary representation, it will have a natural G-invariant inner product induced by complex conjugation and the Haar measure
\begin{equation}
\left\langle \chi_1 \mid \chi_2\right\rangle_{\mathcal{H}_{R}}=\int_G d \mu(g) \chi_1^*(g) \chi_2(g),
\end{equation}
where $\chi_1$ and $\chi_2$ are $G$-equivariant $R$-valued functions on $G$ and $d\mu$ is the Haar measure. Equipped with this inner product,  we can now decompose $L^2\left(\Omega^0(G, R)^G\right)$ into unitary irreducible representations of $G$ which we label as $R$ i.e., the usual orthonormal characters. 

Now having fully characterized the Hilbert spaces, we shift our focus to the Hamiltonian. Interestingly, it turns out that the Hamiltonian for the 2D TrYM theory is the same as 2D topological Yang-Mills theory. We begin with the relevant conjugate field momenta
\begin{equation}
\pi_\theta^a=\frac{1}{e^2}\left(\partial_r A_\theta^a+f^a{ }_{b c} A_r^b A_\theta^c\right),
\end{equation}
using this, it can be seen that the canonical Hamiltonian density has the form
\begin{equation}
\mathscr{H}_C=\ \operatorname{tr}\left[\frac{e^2}{2} \pi_\theta^2-\pi_\theta\left[A_r, A_\theta\right]+\frac{1}{2} T \partial_\theta A_r\right].
\end{equation}
After imposing gauge symmetries \eqref{eqn:GaugeSym} and then using the condition $A_r=0$, we obtain the Hamiltonian density in the following form:
\begin{equation}
\mathscr{H}=\frac{e^2}{2}\int_0^{2 \pi} d \theta  \operatorname{tr}\left(\pi_\theta^2\right)= \frac{1}{2 e^2}\int_0^{2 \pi} d \theta \operatorname{tr}\left(\left(\partial_r A_\theta\right)^2\right),
\end{equation}
which, upon canonical quantization, yields the Laplacian on the group manifold
\begin{equation}
\mathscr{H}=\frac{e^2}{2} \Delta_G, \quad \Delta_G=L_a L_a.
\end{equation}
This Hamiltonian density can be written as the quadratic Casimir operator on the energy eigenstates labeled by an irreducible representation $R$
 \begin{equation}
E=\left\langle\Psi\right| \mathscr{H}\left|\Psi\right\rangle=\frac{e^2}{2} C_2(R) (2\pi).
\end{equation}
\subsection{2D TrYM Partition Function and Holonomies on $\Sigma_{g,b}$} 
\begin{figure}
    \centering
    \includegraphics[width=0.4\linewidth]{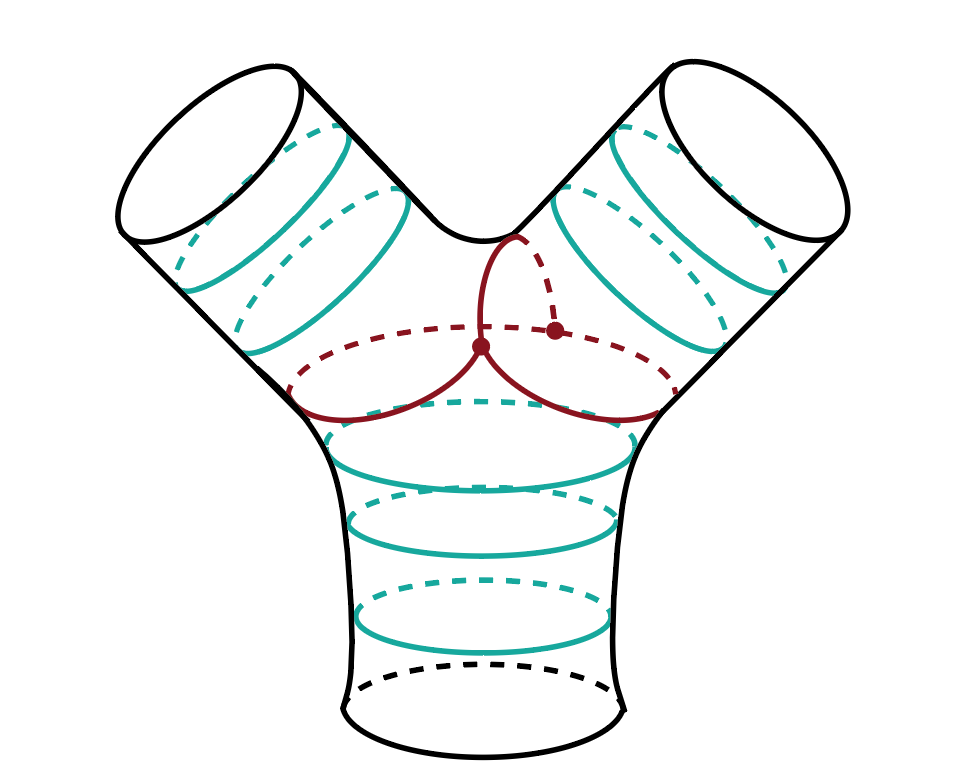}
    \caption{The pair of pants surface. The leaves of the foliation are denoted as teal circles and the singular leaf appears in red. The boundaries $C_i$ are denoted via black circles.}
    \label{fig:fig3}
\end{figure}
Our goal is to investigate what is the effect of inserting tropicalized holonomies along a foliated Riemann surface $\Sigma_{g,b}$ of genus $g$ and $b$ boundaries. To that end, we will first investigate how to first construct the partition function on general closed foliated Riemann surface $\Sigma_{g,0}$ of genus $g>1$ with no boundaries through canonical quantization. It is well known that any such foliated closed Riemann surface of genus $g>1$ admits a handle-body decomposition. Hence, we can construct more complicated amplitudes by gluing together surfaces $\Sigma_{0,3}$  with a pair-of-pants topology, see \figref{fig:fig3}. Let us denote three boundary circles on $\Sigma_{0,3}$ by $C_i$ and let $U_i$ be the corresponding holonomy of the tropical connection $A$ around $C_i$, then the path integral is expected to result in wave functional which resides in the Hilbert space 
\begin{equation}
\mathcal{H=} \bigotimes_{i=1}^3 \mathcal{H}_{C_i}.
\end{equation}
Thus, the corresponding wave functionals $\Psi_3$ must decompose into a linear combination of characters labeled by irreducible representations $R_i$
\begin{equation}
\Psi_3(U_1,U_2,U_3)=\sum_{R_1, R_2, R_3} c_{R_1, R_2, R_3} \prod_{i=1}^3 \chi_{R_i}\left(C_i\right),
\end{equation}
where the coefficients $c_{R_1 R_2 R_3}=\operatorname{dim}\left(\left(R_1 \otimes R_2 \otimes R_3\right)^G\right)$ are precisely the fusion coefficients up to a conventional rescaling. We normalize our Haar measure such that the characters are precisely the orthonormal states, the Peter-Weyl theorem then states that
\begin{equation}
\left\langle R \mid R^{\prime}\right\rangle= \int_G d \mu(g) \overline{\chi_R(g)} \chi_{R^{\prime}}(g)=\delta_{R R^{\prime}},
\end{equation}
which can be used to show
\begin{equation}
N_{R_1 R_2}^{R_3^*} \equiv c_{R_1 R_2 R_3}=\int_G d \mu(g)\chi_{R_1}(g) \chi_{R_2}(g) \overline{\chi_{R_3}(g)}.
\end{equation}

At first glance, the wave functionals might look like an identical copy of what appears in the standard isotropic 2D topological Yang-Mills theory; however, we must recall that our gauge symmetries have been deformed into foliation-preserving gauge symmetries, and consequently, only our angular holonomies that run along the leaves of the foliation contribute
\begin{equation}
U_{\theta}= \operatorname{Tr}\mathcal{P}\exp \left(\int_0^{2 \pi} A_\theta\left(\theta\right) d \theta\right), \quad U_{r}= \operatorname{Tr}\mathcal{P}\exp \left(\int_0^{2 \pi} A_r\left(\theta\right) d \theta\right) = 1.
\end{equation}
As argued in \cite{Albrychiewicz:2025yyg}, for each foliated cylinder, the only free datum is the holonomy along the leaves of the foliation which contributes $\operatorname{rank}(\mathfrak{g})$ degrees of freedom. Any additional holonomies that are evaluated on non contractible cycles along the radial direction $r$, which is transverse to the leaves of the foliation, do not contribute. 

The gluing at the juncture forces the three boundary holonomies to multiply to the identity since a loop around the entire pair of pants is contractible giving the constraint
$$
U_{R_1} U_{R_2} U_{R_3}=1.
$$
In addition to this, gauge invariance at the boundary manifests itself as a fusion constraint, which collapses the sum into a single irreducible representation $R$ i.e.,
\begin{equation}
\Psi_3\left(U_1, U_2\right)=\sum_Rc_R \chi_R\left(U_1\right) \chi_R\left(U_2\right) \chi_R\left(\left(U_1 U_2\right)^{-1}\right),
\end{equation}
we can explicitly see from this that the wave functional is determined by $2\operatorname{rank}\mathfrak{g}$ parameters as expected. What is novel in the tropical analogue, is that we can globally place all holonomies in the same Cartan subalgebra which allows us to commute all holonomies, resulting in
\begin{equation}
\Psi_3\left(U_1, U_2\right)=\sum_R c_R\chi_R\left(U_1\right) \chi_R\left(U_2\right) \chi_R\left(U_1^{-1} U_2^{-1}\right).
\end{equation}
This shows that although the process of ``tropicalization" simplifies the theory, it does not necessarily trivialize the theory. Normalization of the wave functional then enforces
\begin{equation}
\Psi_3\left(U_1, U_2\right)=\sum_R \frac{1}{\sqrt{\operatorname{dim} R}} \chi_R\left(U_1\right) \chi_R\left(U_2\right) \chi_R\left(U_1^{-1}U_2^{-1}\right).
\end{equation}
Likewise, the wave functional $\Psi_2$ on the cylinder $\Sigma_{0,2}$ with two boundaries can be computed via standard arguments, the resulting wave functional has a heat kernel structure
\begin{equation}
\Psi_2\left(U_1, U_2 ; T\right)=\left\langle U_2\right| e^{-T \mathscr{H}}\left|U_1\right\rangle=\sum_R \chi_R\left(U_1\right) \chi_R\left(U_2^{-1}\right) \exp \left[-\pi  e^2 C_2(R)  T\right],
\end{equation}
for some propagation time $T$.  Using $\Psi_2$, we can construct the torus partition function by identifying the boundary holonomies $U_1$ and $U_2$ and summing over all possible holonomies, this results in
\begin{equation}
Z_{\Sigma_{1,0}}=\int_G d \mu(U) \Psi_2(U, U ; T)=\sum_R e^{-\pi e^2 T C_2(R)} \int_G d \mu(U)\left|\chi_R(U)\right|^2=\sum_R e^{-\pi e^2 T C_2(R)}.
\end{equation}
From here, we can see that in the strict topological limit where $g\rightarrow0$, the heat kernel becomes a projector that simply counts the number of gauge-invariant characters
\begin{equation}
\lim _{g^2 T \rightarrow 0} Z_{\Sigma_{1,0}}^{\mathrm{TrYM}}=\sum_R 1.
\end{equation}
This sum is naively divergent since it runs over all irreducible representations $R$ of a compact Lie group and therefore must be regularized.  Given these ingredients, we can now glue these amplitudes together.

Each gluing will remove a factor of $(\operatorname{dim} R)^{-1 / 2}$, and we have $3g-3$ internal gluings given by the radial holonomies that do not contribute; overall, all the pair of pants pieces give a net power of $(g-1)$. Consequently, the partition function of the 2D TrYM is given by
\begin{equation}
Z_{\Sigma_{g,0}}^{\operatorname{TrYM}}=\sum_R(\operatorname{dim} R)^{1-g} \exp \left[-A \frac{e^2}{2} C_2(R)\right],
\end{equation}
where $A$ corresponds to the total area of the surface. Notice that the absolute power of $\operatorname{dim} R$ that appears is precisely half the Euler characteristic $\chi(\Sigma_{g,0})$ of the closed Riemann surface. This differs from the isotropic topological Yang-Mills, where the absolute power is given exactly by the Euler characteristic.  As explained in \cite{Witten:1992xu}, in principle, one should be able to match this partition function to the deformed analytic torsion that was discussed in the previous section \secref{subsec:torsion} in order to obtain an overall numerical factor that would properly count the moduli space of tropicalized flat connections. We leave this as an open question.

From here, we can read off the dimension of the underlying moduli space of tropicalized flat connections by first performing a Poisson resummation. We begin by rewriting the sum as a sum over highest weights. We can label the irreducible representations $R$ by their dominant integral highest weights $\lambda$ as
\begin{equation}
C_2(\lambda)=\frac{1}{2}(\lambda, \lambda+2 \rho), \quad \operatorname{dim} \lambda=\frac{\prod_{\alpha>0}(\lambda+\rho, \alpha)}{\prod_{\alpha>0}(\rho, \alpha)},
\end{equation}
where $\rho$ is the Weyl vector and the product runs over positive roots. We recall that when the weight vector is large, one has the asymptotics behavior $C_2(\lambda) \sim \frac{1}{2}(\lambda, \lambda)$, $\operatorname{dim} \lambda \sim C_{\Delta}\|\lambda\|^p$ for some constant $C_\Delta$ which allows us to asymptotically rewrite the $\operatorname{dim}(R)$ factor as
\begin{equation}
(\operatorname{dim} \lambda)^{1-g} \sim C_{\Delta}^{1-g}\|\lambda\|^{p(1-g)} .
\end{equation}
We can replace the restricted sum over integral highest weights into an unconstrained weight-lattice sum up to a multiplicative factor given by the cardinality of the Weyl group $W$ and then perform the Poisson resummation via the identity
\begin{equation}
\sum_{\lambda \in \Lambda_w} f(\lambda)=\frac{1}{\operatorname{vol} \Lambda_w} \sum_{\mu \in \Lambda_r} \hat{f}(\mu),
\end{equation}
where $\Lambda_w$ and $\Lambda_r$ is the weights and roots lattice respectively. Implementing this Poisson resummation leads to a Gaussian which results in the asymptotic expansion
\begin{equation}
Z_{\Sigma_{g,0}}^{\operatorname{TrYM}}(A) \underset{e^ \rightarrow 0}{\sim} \kappa_g\left(e^2 A\right)^{-\frac{1}{2} (\operatorname{rank}\mathfrak{g})(g-1)}\left[1+O\left(e^{-1 / A}\right)\right],
\end{equation}
where $\kappa_g$ is some constant depending on the genus $g$. From here, one is able to read off the dimension of the moduli space of tropicalized flat connections as the exponent of the area factor as $L^d \sim A^{d / 2}$. This results in the dimension of the moduli space of tropicalized flat connections $\mathcal{M}^{\text {trop }}\left(\Sigma_g,G\right)$ being
\begin{equation}
d=\operatorname{dim} \mathcal{M}^{\mathrm{trop}}\left(\Sigma_g\right)=(g-1)\operatorname{{rank}(\mathfrak{g})}=-\frac{\chi(\Sigma_g)}{2}\operatorname{{rank}(\mathfrak{g})}.
\end{equation}
The tropicalization of flat connections results in a moduli space that is less than half of the usual moduli space of flat connections $\mathcal{M}(\Sigma_g,G)$. Assuming that we are dealing with the case where we only have irreducible flat G-connections on a compact Lie group, the dimension is usually given by 
\begin{equation}
\operatorname{dim} \mathcal{M}\left(\Sigma_g, G\right)=(2 g-2) \operatorname{dim} G=-\chi(\Sigma_g) \operatorname{dim}G.
\end{equation}
The difference is easy to explain from the construction of the TrYM theory i.e., the tropicalization procedure effectively removes all holonomies which lie on contractible or non-contractible cycles that are transverse to the leaves of the foliation, this accounts for the $\frac{1}{2}$ factor. We also obtain an additional topological-like symmetry which allows us to put all holonomies into the same Cartan subalgebra which reduces $\operatorname{dim}(G)$ into $\operatorname{rank}(\mathfrak{g})$.

\textbf{\begin{figure}[H]
    \centering
    \includegraphics[width=0.5\linewidth]{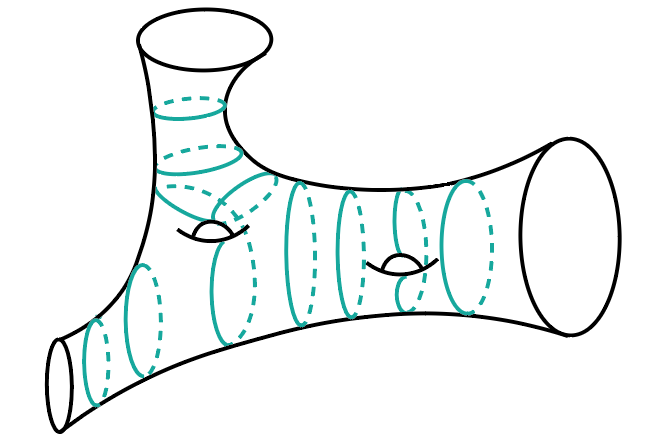}
    \caption{A foliated Riemann surface of genus $g=2$ with $b=3$ boundary components.}
    \label{fig:enter-label1}
\end{figure}}

Finally, we can obtain the heat kernel amplitude $\mathcal{F} _{g,b}$ on Riemann surface of genus $g$ with $b$ boundaries via gluing $b$ cylindrical propagators to our result on a closed Riemann surface
\begin{equation}
\mathcal{F}_{g, b}^{\mathrm{TrYM}}\left(U_1, \ldots, U_b \right)=\sum_R(\operatorname{dim} R)^{1-g-\frac{b}{2}} \chi_R\left(U_1\right) \cdots \chi_R\left(U_b\right) e^{-\frac{1}{2} A e^2 C_2(R)}.
\end{equation}
Once again, we can see that the weighted contribution is half the expected Euler characteristic.

\subsection{Explicit Random Matrix Model for 2D TrYM: The $\beta=1$ GUE.}
We saw in the previous section that the partition function for 2D TrYM was given by
\begin{equation}
Z_{\Sigma_{g,0}}^{\operatorname{TrYM}}=\sum_R(\operatorname{dim} R)^{1-g} \exp \left[-A \frac{e^2}{2} C_2(R)\right],
\end{equation}
from this form, one is able to explicitly see that the primary consequence of this is to have $\frac{1}{2}$ of the usual exponent of the Vandermonde determinant. For example, if we reduce to the case where we are investigating the TrYM on a foliated sphere, then we have
\begin{equation}
Z_{\mathbb{TP}^1}(A)=\sum_R(\operatorname{dim} R) \exp \left(-\frac{Ae^2}{2N} C_2(R)\right).
\end{equation}
We will follow the analysis by \cite{rusakov1990loop, Rusakov:1992uf, Douglas:1993iia} and thus have consequently rescaled our coupling to have an additional $\frac{1}{N}$ factor. If we consider the case of U$(N)$, the sum over irreducible representations $R$ can be characterized by the components of the highest weights which obey
\begin{equation}
n_1>n_2>\cdots>n_N, \quad n_i \in \mathbb{Z} .
\end{equation}
One can then use the standard formulas for the quadratic Casimir
\begin{equation}
C_2(R)=\sum_{i=1}^Nn_i\left(n_i-2i+N+1\right),
\end{equation}
and the Weyl formula 
\begin{equation}
\operatorname{dim} R=\prod_{1 \leq i<j \leq N} \frac{n_i-n_j+j-i}{j-i} .
\end{equation}
to rewrite the partition function (via $k_i:=n_i+N-i+\frac{1}{2}$ with $k_i>k_j$ for $i<j$) up to a constant  as
\begin{equation}
\sum_{k_1>\cdots>k_N}\left[\prod_{i>j}\left(k_j-k_i\right)\right] \exp \left[-\frac{Ae^2}{2N} \sum_{i=1}^N(k_i-N)^2\right] ,
\end{equation}
, this can thus be approximated by an eigenvalue ensemble of a random matrix model
\begin{equation}
Z_{\mathbb{TP}^1}(A)=\int_{\mathbb{R}^N}\prod d\lambda_i\left[\prod_{i>j}\left(\lambda_j-\lambda_i\right)\right] \exp \left[-\frac{Ae^2}{2N} \sum_{i=1}^N(\lambda_i-N)^2\right] ,
\end{equation}
where we have eliminated the $N!$ factorial by integrating over the entirety of $\mathbb{R}^N$ and replaced the discrete indices $k_i$ with continuous eigenvalues $\lambda_i$. It is interesting to point out in hindsight that this calculation suggests that the correct ensemble for the 2D TrYM with unitary group SU$(N)$ results in a new kind of random matrix theory in which we still expect to have Hermitian matrices arising from the fact that we began with a U$(N)$ theory however the Dyson index has now shifted to be that of the standard Gaussian orthogonal ensemble (GOE) of real symmetric matrices.  This result is exactly what one would expect from the intuition that tropicalization reduces complex geometric objects to their real counterparts. Through the realization of this explicit matrix model, one is able to explicitly compute any observable of interest in the 2D TrYM through the usual methods of random matrix theory.

We should note that the techniques of random matrix theory \cite{Eynard:2015aea, Akemann:2015, Livan:2017} cannot be applied blindly due to the fact that there are many standard methods associated to the GUE/GOE that will not directly be usable. This is due to a mixed construction e.g., we should expect to find different bulk local statistics since now the Dyson index $\beta=1$ and hence we should have GOE spacing laws for Hermitian matrices.  We expect that there would be natural consequences of the tropicalization on the edge statistics of the largest eigenvalues and the correlators of the random matrix model. For bookkeeping purposes, we will call this the $\beta=1$ GUE. The standard GUE would therefore be a $\beta=2$ GUE. 

More interestingly from the point of view of physics, one might suspect that having this sort of interpolation between the GUE and GOE might actually give room to expect that the corresponding string dual now has a topological genus expansion that receives contributions from unoriented surfaces like crosscaps \cite{Stanford:2019vob, DiUbaldo:2025cmg}.

\section{Conclusions: Inching Towards Nonequilibrium String Theory}
\label{sec:Conclusions}

In this work we have shown that the tropicalization of 2D topological gauge theories provides a controlled laboratory in which anisotropic scaling and topological invariance coexist in a single exactly-solvable framework. We have provided a more detailed investigation of the moduli space of tropicalized flat connections of the 2D TBF theory through an explicit path integral calculation as well as through the canonical quantization of the theory. 

In performing the path integral quantization, we have shown that one is still able to define a formal analogue of the Hodge star which is induced by the underlying Jordan structure of the TBF theory and used this to construct a Nicolai map which allows the evaluation of the path integral on a cylinder. It is natural to ask whether the methods that were employed here through the usage of the Jordan star can be naturally extended to a Riemann surface of arbitrary genus g through an application of abelianization along the lines of \cite{Blau:1993hj, Blau:1995rs}.

We were also able to show that one is able to explicitly extend the canonical quantization methods of standard isotropic 2D topological Yang-Mills theory to its anisotropic tropical analogue  known as 2D TrYM. In doing so, we have found that the Hilbert space of states is still spanned by equivariant characters in some irreducible representation of the gauge group however the underlying group elements that represent the holonomies were all put into the same Cartan subalgebra due to an additional symmetry which arises upon tropicalization. 

Our analysis was all done in the second order formalism, but an easily answerable and natural question that may arise is how this analysis works in the first order formalism. In the first order formalism, we have additional observables that are given by the $B$ field which is naturally present in the underlying BF model. Unlike the standard 2D BF, we expect that in the 2D TBF theory, these observables would satisfy additional projectability conditions that makes them invariant under translations of points that lie on the leaves of the foliation. The only direction that is effectively seen would be transverse to the leaves of the foliation.

\begin{figure}[H]
    \centering
    \includegraphics[width=0.6\linewidth]{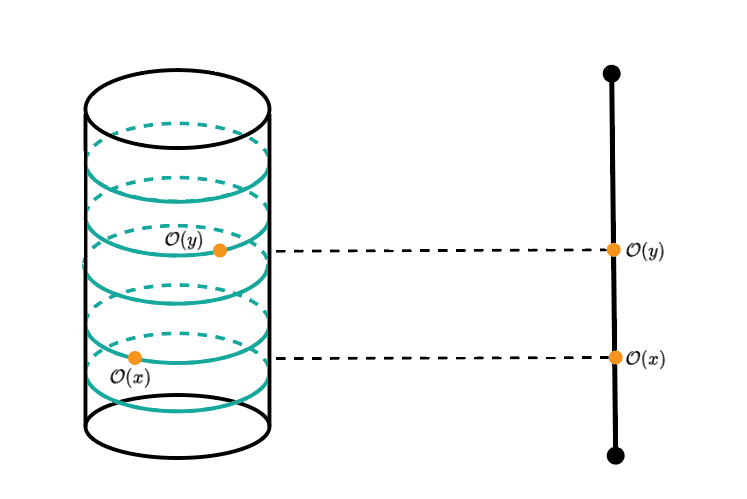}
    \caption{The observables $\mathcal{O}$ would depend on the radial direction transverse to the leaves of the foliation but from the quotient space perspective, it would not dependent on the angular position.}
    \label{fig:label4}
\end{figure}

It has been shown that the first order formulation of standard 2D topological Yang-Mills can lead to a matrix model interpretation through either a lattice regularization or through a direct calculation using Schur polynomials for the case of U$(N)$ gauge groups \cite{Szabo:2013vva}. Regardless of the approach taken, one can see that the discrete Gaussian matrix model that arises acquires a Vandermonde determinant, which is raised to the power of the Euler characteristic of the underlying Riemann surface. We found that a similar formula holds for the tropical analogue except with an usual exponent given by half the Euler characteristic for genus $g>1$, and the matrix model would only get matrix degrees of freedom associated to the surviving holonomies that live along the leaves of the foliation.  We have naturally interpreted this as arising from the usual effects of tropicalization in which complex objects are replaced by their real counterparts. There are many natural followup questions that one can consider. The most obvious one would be to explicitly compute the correlation functions, edge statistics and asymptotic distributions of these generalized random matrix models. Less obvious questions would be to see how tropicalization extends to the case of Gaussian symplectic ensembles with various Dyson indices. One would also like to construct similar generalized random matrix models for other tropological field theories in order to explicitly calculate observables. 

For example, it is well known that one is able to perform q deformations of 2D topological Yang-Mills \cite{Aganagic:2004js}. With the tropical analogue 2D TrYM now constructed and with its easily constructable matrix model, one can naturally pose the categorically induced question, ``what is the q-deformation of 2D TrYM?" Does the q-deformed tropical analogue then have any connections to BPS black holes or the topological string \cite{Aganagic:2004js, Iliesiu:2019lfc} which is constructed by coupling the tropological sigma model of \cite{Albrychiewicz:2023ngk} to anisotropic topological gravity? These q deformations have been useful tools in mapping out the landscape of strongly coupled dualities in supersymmetric field theories \cite{Gadde:2011uv}. 

Besides arbitrary categorical questions, the main physical motivation for investigating the 2D TrYM lies in the observation that tropicalization appears to be a key component of nonequilibrium string perturbation theory. In \cite{neq, ssk, keq}, it was argued that one can probe some general properties that a nonequilibrium string worldsheet theory might possess through the usage of both the Schwinger-Keldysh formalism and the large $N$ expansion of matrix models. It is well known that in the large N limit, the correlators of the underlying matrix model organize into a perturbation theory weighted by the topology of the associated ribbon graphs. Remarkably, it was shown that the usual double perturbation series of nonequilibrium quantum field theories, which decomposes into forward and backward time-evolved branches admits a refinement into a triple perturbation series. In this refinement, one branch corresponds to worldsheets evolving forward in time, another to those evolving backward, and a third, novel branch connects the two: the so-called wedge region. The ribbon diagrams associated with this wedge region are still described by a two-dimensional topological manifold; however, from the perspective of the metric tensor, the geometry is anisotropic.

The role of 2D tropical topological Yang-Mills (TrYM) theory in this picture is to probe the physical properties characteristic of wedge region worldsheet theories. In particular, the anisotropy of the geometry manifests as a projectability condition on the fields. To determine whether 2D TrYM can indeed serve as a wedge region string theory, one should first examine the large $N$ expansion of its associated matrix model and investigate the structure of the resulting perturbative series. We leave this as an open question for future work.

\acknowledgments
We wish to thank both Petr Ho\v{r}ava, Ori Ganor and Jesus Sanchez Jr. for providing useful discussions which stimulated the direction for this short paper. This work has been supported by the Leinweber Institute
for Theoretical Physics.

\bibliographystyle{JHEP}
\bibliography{Bibliography/0DQauntizationTropYM}

\end{document}